\documentclass[preprint,showpacs,preprintnumbers,amsmath,amssymb,
nofootinbib,eqsecnum,12pt,floatfix]{revtex4}

\usepackage{graphicx}
\usepackage{bm}
\usepackage{eufrak}

\newlength{\dummysp}
\newcommand{\half}{\frac{1}{2}}

\newcommand{\beq}{\begin{eqnarray}}
\newcommand{\eeq}{\end{eqnarray}}
\newcommand{\nnn}{ \nonumber \\ }
\newcommand{\p}{{\partial}}

\newcommand{\e}{{\epsilon}}
\newcommand{\s}{{\sigma}}
\newcommand{\vev}[1]{{\langle #1 \rangle}}

\newcommand{\ord}[1]{{{\cal O}(#1)}}
\newcommand{\gappeq}{\mathrel{\rlap {\raise.5ex\hbox{$>$}}
{\lower.5ex\hbox{$\sim$}}}}
\newcommand{\lappeq}{\mathrel{\rlap{\raise.5ex\hbox{$<$}}
{\lower.5ex\hbox{$\sim$}}}}
\newcommand{\myref}[1]{(\ref{#1})}

\newcommand{\ben}{\begin{enumerate}}
\newcommand{\een}{\end{enumerate}}

\newcommand{\sqtw}{\sqrt{2}}

\newcommand{\hc}{{\rm h.c.}}

\newcommand{\ddd}{\nnn &&}

\newcommand{\bit}{\begin{itemize}}
\newcommand{\eit}{\end{itemize}}

\newcommand{\susy}{supersymmetry}

\newcommand{\Ncal}{{\cal N}}

\newcommand{\Ocal}{{\cal O}}

\newcommand{\phit}{{\tilde \phi}}
\newcommand{\phitb}{{\phit^*}}

\newcommand{\chit}{{\tilde \chi}}

\newcommand{\Ftil}{{\tilde F}}
\newcommand{\Ftilb}{\Ftil^*}
\newcommand{\Fcal}{{\cal F}}

\def\[{\left [}
\def\]{\right ]}
\def\({\left (}
\def\){\right )}

\def\nott#1{\setbox0=\hbox{$#1$}                
   \dimen0=\wd0                                 
   \setbox1=\hbox{/} \dimen1=\wd1               
   \ifdim\dimen0>\dimen1                        
      \rlap{\hbox to \dimen0{\hfil/\hfil}}      
      #1                                        
   \else                                        
      \rlap{\hbox to \dimen1{\hfil$#1$\hfil}}   
      /                                         
   \fi}                                         %

\begin{document}

\author{Chen Chen}
\email{chenc10@rpi.edu}
\author{Joel Giedt}
\email{giedtj@rpi.edu}
\author{Joseph Paki}
\email{pakij@rpi.edu}
\affiliation{Department of Physics, Applied Physics and Astronomy,
Rensselaer Polytechnic Institute, 110 8th Street, Troy NY 12065 USA}

\date{\today}

\begin{abstract}
We study supercurrent conservation for the four-dimensional
Wess-Zumino model formulated on the lattice.  The formulation is
one that has been discussed several times, and uses Ginsparg-Wilson
fermions of the overlap (Neuberger) variety, together with an auxiliary
fermion (plus superpartners), such that a lattice version of $U(1)_R$ 
symmetry is exactly preserved in the limit of vanishing bare mass.
We show that the almost naive supercurrent is conserved at one loop.
By contrast we find that this is not true for Wilson fermions
and a canonical scalar action.  We provide nonperturbative
evidence for the nonconservation of the supercurrent in 
Monte Carlo simulations.
\end{abstract}

\title{Supercurrent conservation in the lattice 
Wess-Zumino model with Ginsparg-Wilson fermions}



\maketitle

\section{Introduction}
The formulation of supersymmetric field theories on a spacetime
lattice is of interest because nonperturbative dynamics play an important
role in the theory of supersymmetry breaking and its transmission
to the visible sector of particle physics.  Theories such as
super-QCD and $\Ncal=4$ super-Yang-Mills are of particular
interest, but it seems wise to refine methods using
simpler toy models such as the Wess-Zumino model, give
the difficulties with supersymmetry on the lattice.
Hence, we continue our investigations of a lattice formulation
that was studied by several groups a few years ago \cite{Fujikawa:2001ka,Fujikawa:2001ns,
Fujikawa:2002ic,Bonini:2004pm,Kikukawa:2004dd,Bonini:2005qx}.
What we are developing here is a methodology for
analyzing the extent to which supersymmetry is
a feature of the low energy effective theory.  Since
the lattice formulation explicitly breaks supersymmetry,
this symmetry must be accidental.  In fact, it will
arise from a fine-tuning of bare lattice parameters,
corresponding to the ultraviolet
definition of the theory.  In order to identify the
supersymmetric point in that parameter space, we must
detect the conservation of the supercurrent.  That is a nontrivial task since the naive
supercurrent will mix with other operators, due to
the explicit violation of the symmetry by the discretization.

The problems that we face are not by any means unique to
the Wess-Zumino model.
Four-dimensional supersymmetric models on the lattice\footnote{For reviews 
with extensive references see \cite{Giedt:2006pd,Giedt:2007hz,Giedt:2009yd,
Catterall:2009it,Feo:2004kx,Montvay:2001aj}.} generically
require fine-tuning of counterterms.  This is to be contrasted
with lower dimensional theories where lattice symmetries
can eliminate the need for such fine-tuning; see \cite{Catterall:2009it} for further details.
The one known four-dimensional exception is pure $\Ncal=1$ super-Yang-Mills
using Ginparg-Wilson fermions; the domain wall variety has been
the subject of past \cite{Fleming:2000fa} and recent \cite{Giedt:2008cd,Endres:2008tz,Giedt:2008xm,
Endres:2009yp,Giedt:2009yd,Endres:2009pu} simulations.
Clearly we would like to go beyond pure $\Ncal=1$ super-Yang-Mills,
and in fact all other models contain scalar fields---which are the
source of many difficulties due to unwanted renormalizations that
cannot be forbidden by symmetries.
Recently it was proposed \cite{Elliott:2008jp} that an acceptable amount of fine-tuning
could be efficiently performed using a multicanonical Monte Carlo \cite{Berg:1991cf}
simulation together with Ferrenberg-Swendsen reweighting \cite{Falcioni:1982cz,Ferrenberg:1988yz,Ferrenberg:1989ui}
in a large class of theories; see also \cite{Giedt:2009yd}.
In any such program, it is necessary to study the divergence of the
supercurrent and its renormalization, such as we are doing here
for the Wess-Zumino model.

\subsection{Summary of our previous work}
The theory that we study is the four-dimensional Wess-Zumino model,
formulated on the lattice with overlap (Neuberger) fermions \cite{Neuberger:1997fp},
as well as numerous auxiliary fields.
The goal of the formulation is to impose the Majorana
condition and simultaneously preserve the chiral $U(1)_R$ symmetry \cite{Fujikawa:2001ka,Fujikawa:2001ns,
Fujikawa:2002ic,Bonini:2004pm,Kikukawa:2004dd,Bonini:2005qx}
that is present in the continuum in the massless limit.  As was shown in 
our recent work \cite{Chen:2010uca}, preserving this symmetry significantly limits the number of
counterterms that must be fine-tuned in order to obtain
the supersymmetric continuum limit.  
In addition to overlap fermions, the lattice formulation
has auxiliary fermions (plus superpartner fields) that couple to the overlap fermions
through the Yukawa coupling, as in \cite{Luscher:1998pqa}.  It is possible to integrate
out the auxiliary fermions (and superpartner fields), and when one does this a nonanalytic
Dirac operator results for the surviving fermionic field.
Thus, as has been discussed originally in \cite{Fujikawa:2001ka},
and at greater length in \cite{Fujikawa:2001ns,Kikukawa:2004dd} the action is
singular once auxiliary fields are integrated out.  However,
as we described in \cite{Chen:2010uca}, there is a sensible resolution of
this singularity by taking the theory to ``live'' inside
a finite box, with antiperiodic boundary conditions in the
time direction for the fermions.
The singularity of the Dirac operator that this resolves
is related to nonpropagating modes in the infinite
volume limit; the fact that these are nonpropagating
was shown in \cite{Bonini:2005qx}.  However, singularities
in the Dirac operator raise the spectre of possible nonlocalities
in the continuum limit, as was found in gauge theories with
the SLAC derivative \cite{Karsten:1979wh}.  In \cite{Chen:2010uca} we measured the
degree of localization of the Dirac operator following
the approach of \cite{Hernandez:1998et}.  We found that
while there is localization, it is less pronounced than the
exponential localization of the overlap operator.

The divergences that need to be cancelled in order to
renormalize the lattice theory at one-loop turn out to be strictly wave function renormalization.
The wave function renormalization of the
fermion and the physical scalar match at one loop in the continuum
limit of the lattice expressions; but, the auxiliary
scalar has a mismatched wave function counterterm.  These findings appeared
previously in \cite{Fujikawa:2001ka}; thus our work \cite{Chen:2010uca}
was a confirmation of those results.

\subsection{Plan of this paper}
In Section \ref{defin} we define the Wess-Zumino model both
in the continuum and on the lattice.  We discuss the $U(1)_R$
symmetry of the lattice theory, as well as supersymmetry
transformations that are a symmetry in the limit of the free
theory.  Finally we discuss the fine-tuning action that must
be used to obtain a supersymmetric continuum limit.
In Section \ref{superc} we describe the supercurrent
and an almost naive transcription of it to the lattice.
We also briefly touch on the form of the renormalized
supercurrent in terms of bare lattice operators.
In Section \ref{perta} we perform a one-loop perturbative
analysis of the four-divergence of the supercurrent.
We find that in lattice perturbation theory the supercurrent
is conserved.  We conclude this section by mentioning
two-loop diagrams where we expect that the asymmetric
self-energy of the auxiliary field (at one-loop) will
be important, leading to nonconservation of the lattice
supercurrent.  In Section \ref{nonpa} we describe the
results of our Monte Carlo simulations, where we have
measured the four-divergence of the supercurrent
nonperturbatively.  We find that the violation of
supersymmetry is consistent with contributions beginning
at the two-loop order.  In Section \ref{concl} we
give our concluding remarks.

\section{Definitions}
\label{defin}
\subsection{Continuum}
The Euclidean continuum theory has action
\beq
&& S = - \int d^4 x ~ \bigg\{ \half \chi^T C M \chi + \phi^* \Box \phi + F^* F
+ F^* (m^* \phi^* + g^* \phi^{* 2}) + F (m \phi + g \phi^{ 2}) \bigg\},
\ddd M = \nott{\p} + (m + 2 g \phi) P_+ + (m^* + 2 g^* \phi^*) P_- .
\eeq
Our conventions will be ($i=1,2,3$):
\beq
&& \gamma_0 = \begin{pmatrix} 0 & 1 \cr 1 & 0 \end{pmatrix}, \quad
\gamma_i = \begin{pmatrix} 0 & i\s_i \cr -i\s_i & 0 \end{pmatrix}, \quad
\gamma_5 = \begin{pmatrix} -1 & 0 \cr 0 & 1 \end{pmatrix}, \quad
\ddd P_\pm = \half (1 \pm \gamma_5), \quad C = \gamma_0 \gamma_2 .
\eeq
It can be checked that the action is invariant under
the \susy\ transformations
\beq
&&
\delta_\e \phi = \sqtw \e^T C P_+ \chi, \quad \delta_\e \phi^* = \sqtw \e^T C P_- \chi, \quad
\ddd
\delta_\e \chi = - \sqtw P_+ (\nott{\p} \phi + F) \e - \sqtw P_- (\nott{\p} \phi^* + F^*) \e,
\ddd \delta_\e F = \sqtw \e^T C \nott{\p} P_+ \chi, \quad 
\delta_\e F^* = \sqtw \e^T C \nott{\p} P_- \chi
\eeq

\subsection{Lattice}
We next discuss the lattice action, which is a special case of 
the formulations of \cite{Fujikawa:2001ka,Fujikawa:2001ns}; we
write the lattice action in forms given in
\cite{Bonini:2004pm,Kikukawa:2004dd,Bonini:2005qx}.
For this, a few lattice derivative operators must be introduced.
\beq
A &=& 1 - a D_W, \quad D_W = \half \gamma_\mu ( \p_\mu^* + \p_\mu) + \half a \p_\mu^* \p_\mu
\nnn
D_1 &=& \half \gamma_\mu ( \p_\mu^* + \p_\mu) (A^\dagger A)^{-1/2}
\nnn
D_2 &=& \frac{1}{a} \[ 1 - \( 1 + \half a^2 \p_\mu^* \p_\mu \) (A^\dagger A)^{-1/2} \]
\nnn
D &=& D_1 + D_2 = \frac{1}{a} \( 1 - A (A^\dagger A)^{-1/2} \)
\label{lodf}
\eeq
where $\p_\mu$ and $\p_\mu^*$ are the forward and backward difference
operators respectively.  Note that $D$ is the overlap Dirac
operator.  The lattice action is \cite{Kikukawa:2004dd}:
\beq
S & = & -a^4 \sum_x \bigg\{ \half \chi^T C D \chi
+ \phi^* D_1^2 \phi  + F^* F + F D_2 \phi + F^* D_2 \phi^*
\ddd - \frac{1}{a} X^T C X - \frac{2}{a} \( \Fcal \Phi + \Fcal^* \Phi^* \)
\ddd + \half \chit^T C \( mP_+ + m^* P_- + 2 g \phit P_+ + 2 g^* \phitb P_- \) \chit
\ddd + \Ftilb (m^* \phitb + g^* \phit^{*2}) + \Ftil (m \phit + g \phit^{ 2}) \bigg\}
\label{kisu}
\eeq
As can be seen, the kinetic term for the fermion $\chi$ involves the
overlap Dirac operator $D$.  The use of the other operators $D_1$ and $D_2$
in the scalar part of the action is a departure from what one might
do naively, and is the reason for favorable renormalization of the
action at one-loop.
The tilded fields are the linear combinations
\beq
\phit = \phi + \Phi, \quad \chit = \chi + X, \quad \Ftil = F + \Fcal
\eeq
The fields $\Phi, X, \Fcal$ and their conjugates are auxiliary
fields introduced to allow for a lattice realization of the chiral
$U(1)_R$ symmetry in the $m \to 0$ limit:
\beq
&& \delta \chi = i \alpha \gamma_5 \( 1 - \frac{a}{2} D \) \chi + i \alpha \gamma_5 X,
\quad \delta X = i \alpha \gamma_5 \frac{a}{2} D \chi,
\ddd \delta \phi = -3i \alpha \phi + i \alpha \[  \( 1-\frac{a}{2} D_2 \) \phi
- \frac{a}{2} F^* \] + i \alpha \Phi,
\ddd
\delta \Phi = -3i\alpha \Phi + i \frac{a}{2} \alpha \( D_2 \phi + F^* \)
\ddd
\delta F = 3i\alpha F + i \alpha \[ \( 1 - \frac{a}{2} D_2 \) F - \frac{a}{2} D_1^2 \phi^* \]
+ i \alpha \Fcal
\ddd
\delta \Fcal = 3i\alpha \Fcal + i \frac{a}{2} \alpha \( D_2 F + D_1^2 \phi^* \)
\eeq
which takes a particularly simple form on the tilded variables:
\beq
\delta \chit = i \alpha \gamma_5 \chit, \quad \delta \phit = - 2 i \alpha \phit, 
\quad \delta \Ftil = 4 i \alpha \Ftil
\label{tiltra}
\eeq

We will only need the \susy\ transformations of the tilded fields:
\beq
&&
\delta_\e \phit = \sqtw \e^T C P_+ \chit, \quad \delta_\e \phit^* = \sqtw \e^T C P_- \chit,
\ddd
\delta_\e \chit_\beta = - \sqtw (P_+ (D_1 \phit + \Ftil) \e)_\beta - \sqtw (P_- (D_1 \phit^* + \Ftil^*) \e)_\beta,
\ddd \delta_\e \Ftil = \sqtw \e^T C D_1 P_+ \chit, \quad 
\delta_\e \Ftil^* = \sqtw \e^T C D_1 P_- \chit
\label{sttil}
\eeq
This is not a symmetry of the lattice action for $g \not= 0$,
but is a symmetry in the free case.  Our perturbative analysis
in Section \ref{perta} will identify the corresponding
conserved current.

As noted in \cite{Kikukawa:2004dd}, we can integrate out the auxiliary
fields $X,\Phi,\Fcal$, treating the tilded fields as
constant, to obtain the lattice action:\footnote{Integrating out
an auxiliary fermion to obtain the fermionic part of this action
was previously noted in \cite{Fujikawa:2001ns}.  There it was noted that this relates
the fermionic action to the one of \cite{Luscher:1998pqa} by
a singular field transformation.}
\beq
S& =& -a^4 \sum_x \bigg\{ \half \chit^T C M \chit
- \frac{2}{a} \phitb D_2 \phit  + \Ftilb (1 - \frac{a}{2} D_2)^{-1} \Ftil \ddd
+ \Ftilb (m^* \phitb + g^* \phit^{*2}) + \Ftil (m \phit + g \phit^{ 2}) \bigg\} .
\label{lata}
\eeq
This is the lattice action Eq.~(2.14) of \cite{Bonini:2004pm} with a notation that
interchanges $D_1 \leftrightarrow D_2$,
which is equivalent to Eq.~(2.22) of \cite{Fujikawa:2001ns} for the $k=0$ case,
using the identities\footnote{We thank A.~Feo for explaining this point 
and providing us with a derivation
of these relations.}
\beq
\Gamma_5 = \gamma_5 (1 - \frac{a}{2} D), \quad \Gamma_5^2 = 1 - \frac{a}{2} D_2,
\quad D^\dagger D = \frac{2}{a} D_2 .
\eeq
The fermion matrix is:
\beq
M = \nott{D} + mP_+ + m^* P_- + 2 g \phit P_+ + 2 g^* \phitb P_- ,
\quad \nott{D} = (1 - \frac{a}{2} D_2)^{-1} D_1
\label{fmat}
\eeq
This way of writing the Dirac operator can be related to the
one that appears in \cite{Fujikawa:2001ns} by the identity:
\beq
(1 - \frac{a}{2} D_2)^{-1} D_1 = (1 - \frac{a}{2} D)^{-1} D
\eeq
Furthermore we can integrate out the auxiliary
fields $\Ftil, \Ftilb$ to obtain the action
\beq
S& =& a^4 \sum_x \bigg\{ -\half \chit^T C M \chit
+ \frac{2}{a} \phitb D_2 \phit  
+ (m^* \phitb + g^* \phit^{*2}) (1 - \frac{a}{2} D_2) (m \phit + g \phit^{ 2}) \bigg\}
\label{acws}
\eeq
This is the action that is used in our numerical simulations.

When fine-tuning of the lattice action is performed,
we must invoke the most general lattice action consistent
with symmetries.  Since we perform our simulations at $m \not= 0$,
this is just the action with all dimension $\leq 4$ operators
built out of the physical fields, $\phit$ and $\chit$.
We write it here for reference:
\beq
S &=&  a^4 \sum_x \bigg\{ -\half \chit^T C ( \nott{D} + m_1 P_+ + m_1^* P_- ) \chit
+ \frac{2}{a} \phitb D_2 \phit  
\ddd
+ m_2^2 |\phit|^2 + \lambda_{1} |\phit|^4
+ \big( m_3^2 \phit^2 + g_1 \phit^3 
+ g_2 \phit \phit^{*2} + \lambda_2 \phit^4 + \lambda_{3} \phit \phit^{*3} + \hc \big)
\ddd
- \chit^T C ( y_1 \phit P_+ + y_1^* \phit^* P_- ) \chit
- \chit^T C ( y_2 \phit P_- + y_2^* \phit^* P_+ ) \chit \bigg\}
\label{mogeac}
\eeq
A term linear in $\phit$ has been eliminated
by the redefinition $\phit \to \phit + c$ with $c$ a constant.  The parameters
$m_2^2$ and $\lambda_1$ are real and all other parameters
are complex.  Whereas in
the supersymmetric theory there are four real parameters,
in the most general theory we have eighteen real parameters to
adjust.  Holding $m_1$ and $y_1$ fixed (corresponding
to some choice of values for $m_R$ and $g_R$ in the
long-distance effective theory), we have fourteen 
real parameters that must be adjusted to obtain the supersymmetric
limit.  The counting can be alleviated somewhat by imposing
CP invariance, so that all parameters can be assumed real.
Then we have a total of ten parameters.  Holding two fixed,
we must tune the other eight to achieve the supersymmetric limit.
Conducting a fine-tuning in an eight-dimensional parameter
space is a daunting task.

On the other hand in the limit $m_1 \to 0$ we can impose the
$U(1)_R$ symmetry \myref{tiltra}.  This restricts the action
to
\beq
S &=&  a^4 \sum_x \bigg\{ -\half \chit^T C \nott{D} \chit
+ \frac{2}{a} \phitb D_2 \phit  
+ m_2^2 |\phit|^2 + \lambda_{1} |\phit|^4
\ddd
- \chit^T C ( y_1 \phit P_+ + y_1^* \phit^* P_- ) \chit
\bigg\}
\label{symgeac}
\eeq
If we hold $y_1$ fixed (corresponding to some
choice of $g_R$ in the long-distance effective
theory), then only $m_2^2$ and $\lambda_1$ must be
fine-tuned.  Conducting a search in a two-dimensional parameter
space, with both coming from bosonic terms, is manageable.
The difficult part is that we must extrapolate to the massless
fermion limit.  Another potential problem is that we impose
antiperiodic boundary conditions for the fermion in the time
direction, but must impose periodic boundary conditions for
the scalar in order for the action to be single-valued 
on the circle in the time direction.  This breaks supersymmetry explicitly by boundary
conditions.  At finite mass this should be an effect that
can be made arbitrarily small by taking the large volume limit.
However at vanishing mass, there will be long distance modes
that will ``feel'' the breaking due to boundary conditions.\footnote{We
thank G.~Bergner for raising this point.}  Thus it is important
that we take $T \gg 1/ma$ as $m$ is sent to zero, where $T$ is
the number of sites in the time direction.

What we have seen in \cite{Chen:2010uca} is that the one-loop behavior of
the theory \myref{lata} closely follows that of the continuum, so that no new
operators are generated at this order.  Thus at this level
of approximation, a fine-tuning of the general lattice action \myref{mogeac} is not needed.
Due to this good one-loop behavior it is of interest to study the original lattice
action \myref{acws} in our simulations, without any fine-tuning.  By
measuring the degree of \susy\ breaking through nonconservation
of the supercurrent, we gain
information about the higher orders and nonperturbative
aspects of the lattice theory.

\section{Supercurrent, mixing and renormalization}
\label{superc}
For a general superpotential $W(\phi)$, the supercurrent is
\beq
S^\mu = \sqtw \[ \nott{\p} \phi \gamma^\mu P_- \chi
+ \nott{\p} \phi^* \gamma^\mu P_+ \chi
+ \frac{\p W}{\p \phi} \gamma^\mu P_+ \chi 
+ \(\frac{\p W}{\p \phi}\)^* \gamma^\mu P_- \chi \]
\label{succ}
\eeq
and in our case $\p W/ \p \phi = m \phi + g \phi^2$.
There is also a form with the auxiliary field:
\beq
S^\mu = \sqtw \[ \nott{\p} \phi \gamma^\mu P_- \chi
+ \nott{\p} \phi^* \gamma^\mu P_+ \chi
- F^* \gamma^\mu P_+ \chi 
- F \gamma^\mu P_- \chi \]
\label{sucf}
\eeq
Because of the supersymmetry breaking on the lattice,
this will mix with other operators in the same symmetry
channel.  
If the lattice action (Eq.~\myref{mogeac} in the massive
case and Eq.~\myref{symgeac} in the massless case) is fine-tuned, then in the long
distance effective theory there will be a supercurrent
that is conserved in the continuum limit.  
The way to detect the existence of \susy\ in
this fine-tuning process is to consider linear combinations
of bare lattice operators and search for one
that has vanishing four-divergence in the supersymmetric
limit, modulo contact terms.  We briefly describe that approach
in Section \ref{sufo} below.  Before doing so, we
give an almost naive discretization of the continuum
supercurrent \myref{succ} that will turn out to be
the conserved supercurrent of the one-loop analysis below.

\subsection{Almost naive lattice supercurrent}
We formulate this lattice supercurrent in terms of tilded fields:
\beq
S_\mu &=& \sqtw \[ D_1 \phit \gamma_\mu P_- \chit
+ D_1 \phit^* \gamma_\mu P_+ \chit
+ \frac{\p W}{\p \phit} \gamma_\mu P_+ \chit 
+ \(\frac{\p W}{\p \phit}\)^* \gamma_\mu P_- \chit \] \nnn
\frac{\p W}{\p \phit} &=& m \phit + g \phit^2
\label{latsuc}
\eeq
It is almost naive
because $\nott{\p}$ has been replaced by $D_1$,
rather than a more naive prescription such
as
\beq
D_S = \sum_\mu \gamma_\mu \p_\mu^S, \quad \p_\mu^S = \half (\p_\mu + \p_\mu^*)
\label{symds}
\eeq
Note that \myref{latsuc} is the form of the supercurrent with the
auxiliary fields eliminated.  Thus in working with it we
should use the Feynman rules corresponding to the action
without auxiliary fields.  In our perturbative calculations
below, that will be the action \myref{acws}.  We do not need
to use the more general action \myref{mogeac} for the $\ord{g}$
calculations that we do, since the leading violation of
supersymmetry is an $\ord{g^2}$ nonsupersymmetric
wavefunction renormalization for the auxiliary field,
as will be discussed below.

We could also work with a supercurrent containing
auxiliary fields:
\beq
S^\mu &=& \sqtw \[ D_1 \phit \gamma^\mu P_- \chit
+ D_1 \phit^* \gamma^\mu P_+ \chit
- \Ftil^* \gamma^\mu P_+ \chit 
- \Ftil  \gamma^\mu P_- \chit \] 
\label{latsud}
\eeq
At finite $a$, this will lead to slightly different results than
\myref{latsuc}, since the lattice equations of motion for the
auxiliary field are
\beq
-\Ftil^* = \( 1-\frac{a}{2} D_2 \) ( m \phit + g \phit^2 )
\eeq
In fact we will also consider the form \myref{latsud} in
our perturbative calculations below.  It is a convenient
choice, since we know crucial facts about the nonsupersymmetric
renormalization of the auxiliary field from our previous
one-loop analysis of counterterms.

The variation of the action under the lattice
version of the supersymmetry transformation 
was given in \cite{Kikukawa:2004dd} and is equal to:\footnote{The
variation of the action under a modified supersymmetry transformation
was also given in \cite{Bonini:2004pm}, which reduces to the
result \myref{niac} in the appropriate limits.}
\beq
\delta S = \sqtw a^4 \sum_x \chit^T C \[g P_+ (2 \phit D_1 \phit - D_1 \phit^2) 
+ g^* P_- (2 \phit^* D_1 \phit^* - D_1 \phit^{*2}) \] \e
\label{niac}
\eeq
Note that this is $\ord{a}$, since
\beq
\lim_{a \to 0} \sum_x \chit^T C P_+ (2 \phit D_1 \phit - D_1 \phit^2) = 0
\eeq
Also note the presence of $g$ in \myref{niac}.
In order to see violation of the supersymmetric identity $\p_\mu S_\mu=0$
in the continuum limit, diagrams involving the coupling $g$ must be included.
Loop diagrams are required, in order to get the divergences that
cancel the $\ord{a}$ factor coming from \myref{niac} in the continuum limit.

\subsection{Renormalized supercurrent}
\label{sufo}
Having written down the almost naive supercurrent,
which will be the subject of all of our computations in this paper,
we next mention the more general case.  That is,
the form of the renormalized supercurrent, which is
expected to be different from the almost naive version.
At a given engineering dimension,
we write down all operators that have the index
structure of $S_{\mu \alpha}(x)$.  We denote these
as $\Ocal_{\mu \alpha, j}^{(n/2)}$ where $n/2$ is the engineering dimension;
thus $n$ takes odd values $3,5,7,\ldots$, and the index $j$ labels the
different operators of dimension $n/2$.
A linear combination of these is the long distance effective supercurrent 
at lattice spacing $a$:
\beq
S_{\mu \alpha}(x) = \sum_{n=3,5,7,\ldots} 
\sum_j b_j^{(n/2)} a^{(n-7)/2} \Ocal_{\mu \alpha, j}^{(n/2)}(x)
\label{fpas}
\eeq
Clearly it will be a demanding task to identify this $S_{\mu}$
nonperturbatively.  However, there is no other way to properly
fine-tune the action.  We must find $S_\mu$ and the point
in parameter space where it is conserved.  Before tacking
this problem (in future work), we will (in this paper) examine
the properties of the almost naive supercurrent.

\section{Perturbative analysis}
\label{perta}
Here we compute $\vev{\p_\mu S_\mu(x) \Ocal(0)}$ for a few different
choices of $\Ocal$, with our first choice being $\Ocal=\chit$,
which gets an $\ord{g}$ contribution at one-loop.
As will be seen, this leads naturally to a second
choice, $\Ocal = \phit^* \chit$, which gets an
$\ord{g^0}$ contribution at one-loop.  

All other choices correspond to higher loop diagrams.
As summarized above, we know that $Z_\phi=Z_\chi \not= Z_F$ at
one-loop.  These self-energy diagrams can occur
on an internal line of $\vev{\p_\mu S_\mu(x) \Ocal(0)}$,
giving for example an $\ord{g^2}$ contribution when
$\Ocal = \phit^* \chit$, at two-loops.  Because of the
mismatch of the $Z$ factors, we can be confident
that the self-energies as a function of loop momentum
are also mismatched, so that the cancellations
that occur in the continuum theory will not happen.
An example diagrams will be presented below.  However,
the point is that we will need to be able to
compute two-loop diagrams in order to begin
fine-tuning the action using the conservation
of the supercurrent as a probe.  This is beyond
the scope of the present paper and is left to
future work.

In the perturbative analysis, the following lattice
propagators are used:
\beq
\sum_x a^4 e^{i p \cdot x} \vev{ \chit(x) \chit^T(0) C} & = &
\frac{-D_1(p) + (1 - \frac{a}{2} D_2(p)) (m^* P_+ + m P_-)}
{\frac{2}{a} D_2(p) + (1 - \frac{a}{2} D_2(p)) |m|^2} \nnn
\sum_x a^4 e^{i p \cdot x} \vev{ \phit(x) \phit^*(0) } & = &
\frac{-1}
{\frac{2}{a} D_2(p) + (1 - \frac{a}{2} D_2(p)) |m|^2}
\eeq
where
\beq
D_1(p) = \frac{-i a^{-1} \sum_\mu \gamma_\mu \sin(p_\mu a)}{
\sqrt{ [ 1 - 2 \sum_\mu \sin^2(p_\mu a/2 )]^2  + \sum_\mu \sin^2(p_\mu a)}}
\label{d1mom}
\eeq
\beq
D_2(p) = \frac{1}{a} \( 1 - \frac{ 1 - 2 \sum_\mu \sin^2(p_\mu a/2 )}{
\sqrt{ [ 1 - 2 \sum_\mu \sin^2(p_\mu a/2 )]^2  + \sum_\mu \sin^2(p_\mu a)}} \)
\eeq
The fermion propagator will be represented by a solid
line and the scalar propagator by a dashed line.

\subsection{First choice:  $\Ocal= \chit$}
\begin{figure}
\begin{center}
\includegraphics[width=3in,height=2in,bb= 50 550 350 750,clip]{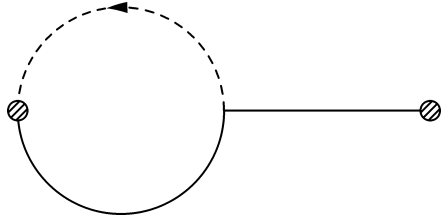}
\caption{The Feynman diagrams for the correlation function that we study.
In actuality, there are two such diagrams, one with operator $D_1 \phit \gamma_\mu P_- \chit$ on
the left-hand side, and one with operator $\phit \gamma^\mu P_+ \chit$ 
on the left-hand side. \label{dfws} }
\end{center}
\end{figure}

It is convenient to work instead with the Fourier transform:
\beq
\sum_x e^{i p \cdot x} \vev{\p_\mu^S S_\mu(x) \chit(0)}
\eeq
The corresponding diagram at one-loop is Fig.~\ref{dfws}.
Note that this includes the point $x=0$ so that we have to worry 
about contact terms.  These would involve the supersymmetric
variation of $\chit$: 
\beq
\vev{\Delta \chit(x)} \delta_{x,0} \propto \vev{P_+(D_1 \phit + \Ftil)(x) 
+ P_-(D_1 \phit^* + \Ftil^*)} \delta_{x,0}
\eeq
At one-loop we found in our previous work $\vev{\Ftil(x)}=\vev{\phit(x)}=0$.
So we do not have to worry about contact terms at one-loop, due
to the absence of tadpoles.

It is a straightforward calculation to obtain the four correlation
functions that we need:
\beq
&& \sum_x e^{i p \cdot x} \vev{\p_\mu^S (D_1 \phit \gamma_\mu P_- \chit)_\alpha(x) \chit_\beta(0)} 
\ddd = -g^* \frac{1}{a} ( e^{-i p_\mu a} - e^{i p_\mu a} ) 
\int \frac{d^4 q}{(2\pi)^4} (D_1(q) 
\gamma_\mu P_- S(p-q) P_- S(p) C)_{\alpha \beta} G(q) \\
&& \sum_x e^{i p \cdot x} \vev{\p_\mu^S (D_1 \phit^* \gamma_\mu P_+ \chit)_\alpha(x) \chit_\beta(0)}
\ddd = - g \frac{1}{a} ( e^{-i p_\mu a} - e^{i p_\mu a} ) 
\int \frac{d^4 q}{(2\pi)^4} (D_1(q) 
\gamma_\mu P_+ S(p-q) P_+ S(p) C)_{\alpha \beta} G(q) \\
&& \sum_x e^{i p \cdot x} \vev{\p_\mu^S (\phit \gamma_\mu P_+ \chit)_\alpha(x) \chit_\beta(0)}
\ddd = - g^* \frac{1}{a} ( e^{-i p_\mu a} - e^{i p_\mu a} ) 
\int \frac{d^4 q}{(2\pi)^4} (\gamma_\mu P_+ S(p-q) P_- S(p) C)_{\alpha \beta} G(q) \\
&& \sum_x e^{i p \cdot x} \vev{\p_\mu^S (\phit^* \gamma_\mu P_- \chit)_\alpha(x) \chit_\beta(0)}
\ddd = - g \frac{1}{a} ( e^{-i p_\mu a} - e^{i p_\mu a} ) 
\int \frac{d^4 q}{(2\pi)^4} (\gamma_\mu P_- S(p-q) P_+ S(p) C)_{\alpha \beta} G(q)
\eeq
Notice that these correlation functions have certain common factors.  On the left,
\beq
\frac{1}{2a} ( e^{-i p_\mu a} - e^{i p_\mu a} )
\eeq
and on the right
\beq
S(p) C
\eeq
Thus when we sum them, we can factor out these bits and the remainder
has to be the thing that cancels.
We now check whether the
four-divergence of the almost naive supercurrent vanishes.  We have computed the
above integrals numerically for various values of $a$, having in mind the $a \to 0$ extrapolation.

If the naive supercurrent is to work, then with $p=(p_0,0,0,0)$
\beq
{\mathfrak S(p_0)}_{-,\alpha \beta} &=& 
b_1^{(7/2)} {\mathfrak I(p_0)}_{0-,\alpha \beta} 
+ b_1^{(5/2)} a^{-1} {\mathfrak I(p_0)}_{1-,\alpha \beta}
\nnn
{\mathfrak I(p_0)}_{1-,\alpha \beta} &=& \int \frac{d^4 q}{(2\pi)^4}  
(D_1(q) \gamma_0 P_- S(p-q) P_-)_{\alpha \beta} G(q) 
\nnn
{\mathfrak I(p_0)}_{2-,\alpha \beta} &=& \int \frac{d^4 q}{(2\pi)^4}  
(\gamma_0 P_+ S(p-q) P_-)_{\alpha \beta} G(q)
\label{intwc}
\eeq
the quantity ${\mathfrak S(p_0)}_{-,\alpha \beta}$ has to vanish in the continuum limit.  
It is not hard to show that a naive
continuum limit of this expression does vanish, provided
\beq
b_1^{(7/2)} = \sqtw, \quad b_1^{(7/2)} = \sqtw m a
\label{tteq}
\eeq
i.e., for the proper coefficients of the almost
naive supercurrent.
If ${\mathfrak S(p_0)}_{-,\alpha \beta}$ vanishes in
the one-loop continuum limit, it would also follow that
\beq
{\mathfrak S(p_0)}_{+,\alpha \beta} &=& \int \frac{d^4 q}{(2\pi)^4} \bigg\{ 
b_1^{(7/2)*} (D_1(q) \gamma_0 P_+ S(p-q) P_+)_{\alpha \beta} G(q) 
\ddd + b_1^{(5/2)*} a^{-1} (\gamma_0 P_- S(p-q) P_+)_{\alpha \beta} G(q) \bigg\}
\eeq
would vanish as well.

\begin{table}
\begin{tabular}{|c|c|c|c|} \hline
$a$  & $p_0$ & $(2\pi)^4{\mathfrak I(p_0)}_{1-,0 0}$ 
  & $(2\pi)^4{\mathfrak I(p_0)}_{2-,0 0}$ \\ \hline
0.1 & 0.1 & 2.0375(2) & -2.0323(2) \\
0.1 & 0.2 & 4.0700(4) & -4.0595(4) \\ \hline
0.01 & 0.1 & 4.2796(4) & -4.2789(4) \\
0.01 & 0.2 & 8.5544(9) & -8.5529(9) \\ \hline
0.001 & 0.1 & 6.5529(7) & -6.5513(7) \\
0.001 & 0.2 & 13.1009(13) & -13.0979(13) \\ \hline
\end{tabular}
\caption{Values of the integrals for $m=1$, at various values of
$p_0$ and $a$. \label{svoi} }
\end{table}

When one looks at the specific values of the integrals,
(Table \ref{svoi}) what one finds is that we should set
the coefficients to the naive values of \myref{tteq}.
(Of course
the overall normalization of $\sqtw$ is a matter of convention.)  
That is, the almost naive supercurrent is conserved at one loop.

This result suggests that
at one-loop there is an argument that the quantity we are
computing is determined by the free theory and the supercurrent
is not renormalized at one-loop.  

\subsection{Second choice:  $\Ocal=\phit^* \chit$}
\begin{figure}
\begin{center}
\includegraphics[width=3in,height=2in,bb=50 500 400 775,clip]{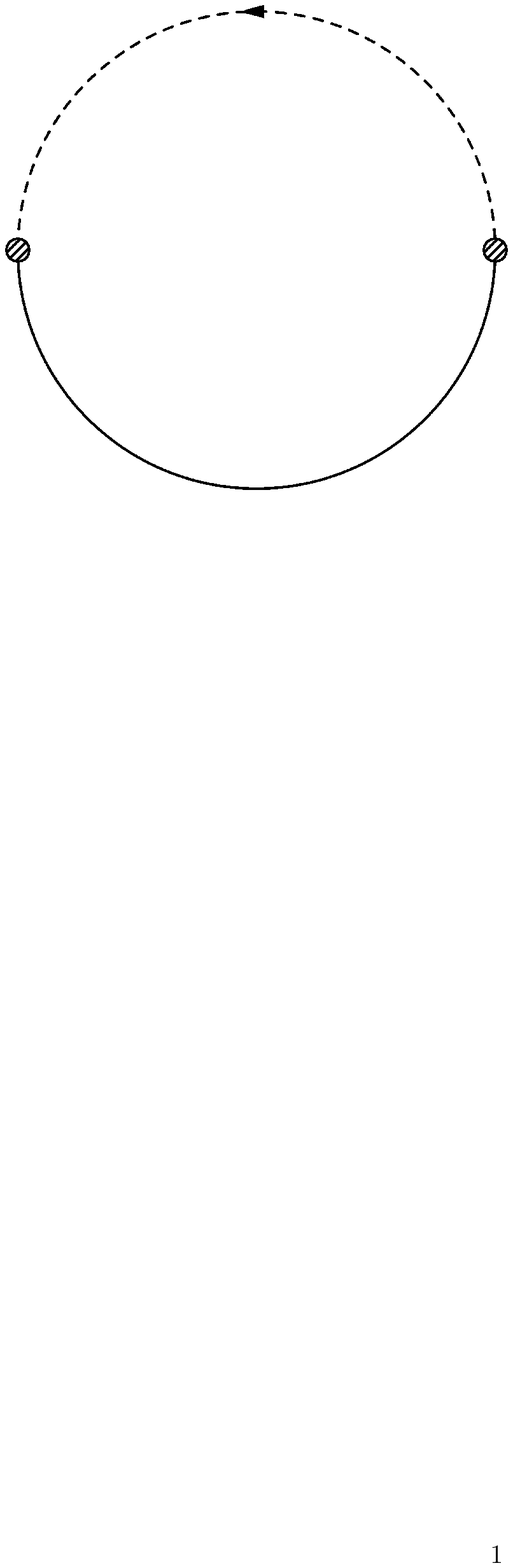}
\caption{The simpler correlation function that must vanish in
the continuum limit. \label{scf1} }
\end{center}
\end{figure}

That the forgoing claim is true can be seen from
the fact that when we amputate the propagator $S(p)$ from
the diagram Fig.~\ref{dfws} (i.e., the fermion propagator that is not in
the loop), we obtain the diagram of Fig.~\ref{scf1}.  However, the latter
diagram is just the one that we would obtain in the
free theory from evaluating $\vev{\partial_\mu^S S_\mu(x) (\phi^* \chi)(0)}$.
Since in the free theory the variation of the action
under supersymmetry is zero, there should be a conserved
current.  What we have just found is that the almost naive
lattice supercurrent is that supercurrent to within
the accuracy of our numerical evaluation of the integrals.

\subsection{A more naive discretization}
The behavior we have just observed is to be contrasted with what happens in
the free theory if we use Wilson fermions and naive scalars so that the action is
\beq
S = - \sum_x a^4 \{ \half \chi^T C (D_W + m) \chi + \phi^* \p_\mu \p_\mu^* \phi
+ F^* F + m ( F \phi + F^* \phi^* ) \}
\label{nfa}
\eeq
where $D_W$ is defined in Eq.~\myref{lodf} and we have specialized
to a real mass $m$.  
Also, we will use the symmetric difference Dirac operator \myref{symds} in the
discretization of the supercurrent \myref{succ}, $\nott{\p} \to D_S$,
yielding 
\beq
D_S(p) = -\frac{i}{a} \sum_\mu \gamma_\mu \sin (p_\mu a)
\eeq
in momentum space.  Corresponding to \myref{nfa}, after
integrating out the auxiliary field, there
will be fermion propagator 
\beq
S_W(p) = \frac{-D_S(p) + m + \frac{2}{a} \sum_\mu \sin^2 (p_\mu a / 2)}
{\frac{1}{a^2} \sum_\mu \sin^2(p_\mu a) + (m + \frac{2}{a} \sum_\mu \sin^2 (p_\mu a / 2))^2}
\eeq
and scalar propagator 
\beq
G_N(p) = \frac{-1}{\frac{4}{a^2} \sum_\mu \sin^2 (p_\mu a / 2) + m^2}
\eeq
It is easy to check that
the action \myref{nfa} is not invariant under a lattice supersymmetry
transformation, such as
\beq
&&
\delta_\e \phi = \sqtw \e^T C P_+ \chi, \quad \delta_\e \phi^* = \sqtw \e^T C P_- \chi,
\ddd
\delta_\e \chi = - \sqtw P_+ (D_S \phi + F) \e - \sqtw P_- (D_S \phi^* + F^*) \e,
\ddd \delta_\e F = \sqtw \e^T C D_S P_+ \chi, \quad 
\delta_\e F^* = \sqtw \e^T C D_S P_- \chi
\label{nstr}
\eeq
in spite of the fact that this is a free theory.  I.e., it
does not have the behavior \myref{niac} seen in the discretization
that is the main focus of this paper.

The integrals that are the counterparts of \myref{intwc} are:
\beq
{\mathfrak I(p_0)}_{3-,\alpha \beta} &=& \int \frac{d^4 q}{(2\pi)^4}  
(D_S(q) \gamma_0 P_- S_W(p-q) P_-)_{\alpha \beta} G_N(q) 
\nnn
{\mathfrak I(p_0)}_{4-,\alpha \beta} &=& \int \frac{d^4 q}{(2\pi)^4}  
(\gamma_0 P_+ S_W(p-q) P_-)_{\alpha \beta} G_N(q)
\eeq
It can be seen from Table \ref{svoj} that setting
\myref{tteq} will not work:  there is a significant
lattice artifact and the naive supercurrent is not
conserved.  This shows the benefits of
the formulation which is the main focus of this
article.

\begin{table}
\begin{tabular}{|c|c|c|c|} \hline
$a$  & $p_0$ & $(2\pi)^4{\mathfrak I(p_0)}_{3-,0 0}$ 
  & $(2\pi)^4{\mathfrak I(p_0)}_{4-,0 0}$ \\ \hline
0.1 & 0.1 & 7.772923 & -2.076717 \\
0.1 & 0.2 & 15.540440 & -4.148735 \\ \hline
0.01 & 0.1 & 65.802(7) & -4.4650(4) \\
0.01 & 0.2 & 131.599(13) & -8.9252(9) \\ \hline
0.001 & 0.1 & 623.85116 & -6.77184 \\
0.001 & 0.2 & 1247.69800 & -13.53882 \\ \hline
\end{tabular}
\caption{Values of the more naive integrals for $m=1$, at various values of
$p_0$ and $a$. \label{svoj} }
\end{table}

\subsection{Higher dimensional operators}
Another interesting thing we have looked at is
\beq
\int \frac{d^4 q}{(2\pi)^4} 
a (D_1 (q) D_1(q) \gamma_0 P_- S(p-q) P_-)_{\alpha \beta} G(q) 
\eeq
which corresponds to one of the dimension 9/2 operators
that would appear in \myref{fpas}.  The power of $a$ could possibly
be overcome by a $1/a$ coming from the extra $D_1(q)$.
In fact what we find is that the integral is linearly
divergent.  This indicates that higher dimensional operators
could play an important role in the renormalized supercurrent
when we go to higher orders.  Obviously we would prefer
to avoid this possibility, since there is an infinite
number of such operators.  However, we do not at this
stage see a reason for excluding them from the sum \myref{fpas}.

\subsection{Beyond one-loop}
\begin{figure}
\begin{center}
\includegraphics[width=3in,height=2in,bb=75 500 325 700,clip]{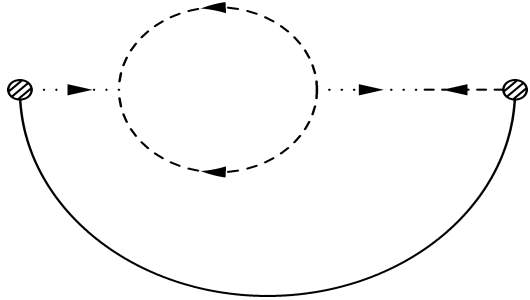}
\caption{A diagram where the nonsupersymmetric renormalization of
the auxiliary field will play a role. The dashed/dotted line
is the mixed $\vev{F \phi}$ propagator.  The dotted line
is the $\vev{F F^*}$ propagator.  \label{dwam} }
\end{center}
\end{figure}

The result of the one-loop investigation is that there is a conserved
supercurrent in the continuum limit, when we work to $\ord{g}$.
This agrees with the fact that the renormalization at
one loop only yields a nonuniform wavefunction renormalization
for the auxiliary field: $Z_F \not= Z_\phi=Z_\chi$ (in the
continuum limit).  This cannot
affect correlation functions at $\ord{g}$ since
the wavefunction renormalization of the auxiliary field
would have to appear on the internal line of a diagram,
which necessarily implies a factor of $g^2$ to occur.
For example, the diagram in Fig.~\ref{dwam} would be sensitive
to the value of $Z_F$, and is $\ord{g^2}$.  Note that
for this analysis we have switched to the supercurrent
involving the auxiliary field, Eq.~\myref{sucf}.
In the nonperturbative analysis that we discuss next,
we find further evidence that the nonconservation
of the almost naive supercurrent begins at $\ord{g^2}$.

\section{Nonperturbative analysis}
\label{nonpa}
Here it is convenient to take the spatial transform
so that we instead work with 
\beq
a^3 \sum_{\bm x} ~ \p_\mu^S S_{\mu \alpha}(t, \bm x) 
= a^3 \sum_{\bm x} ~ \p_t^S S_{0 \alpha}(t, \bm x)
= \p_t^S Q_\alpha(t) .
\eeq
We evaluate 
\beq
C(t) = \sum_{{\bf x}} \vev{ \p_\mu^S S_\mu(t,{\bf x}) \Ocal(0) }
\eeq
Since $S_{\mu \alpha}$ is fermionic, an odd number of $\chit$ fields must appear
in nonvanishing correlation functions with $\p_t Q_\alpha(t)$.  The case
that we will consider is $\Ocal = \chit^T C$.
Because of cluster decomposition, $C(t)$ will fall off with $t$
exponentially, governed by the mass $m_{\text{eff}}$ of the lightest state
created by $\Ocal$.  As we tune the action, $m_{\text{eff}}$ and
the fall-off with $t$ will change.  Thus we could mistake
a decrease in $C(t)$ for an improvement of \susy\ when it is
really an increase in $m_{\text{eff}}$.  Similarly, we could think
we have worsened \susy\ when in fact all we did was to decrease
$m_{\text{eff}}$.  Clearly we need a way to normalize $C(t)$ in order
to cancel off this $\exp(-t ~ m_{\text{eff}})$ behavior.  For this reason
we look instead at the ratio
\beq
R(t) = \frac{ | \sum_{{\bf x}} \vev{ \p_\mu S_{\mu \alpha}(t,{\bf x}) (\chit^T C)_\beta(0) } |}
{| \sum_{{\bf x}} \vev{ \chit_\alpha(t,{\bf x}) (\chit^T C)_\beta(0) } |}
\label{ratwm}
\eeq
and will set $\alpha=\beta=0$.

\begin{figure}
\begin{center}
\includegraphics[width=3in,height=5in,angle=90]{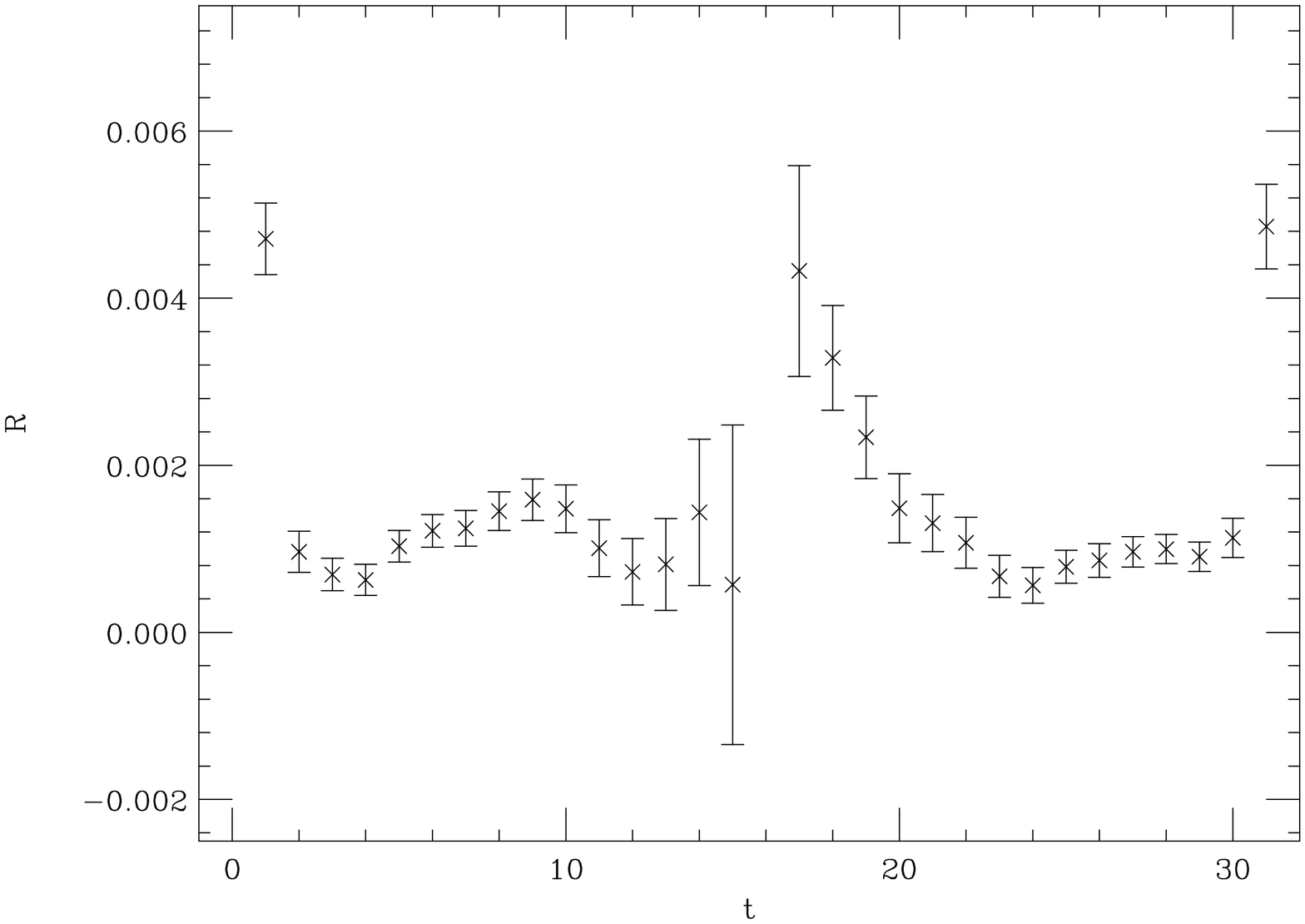}
\caption{The ratio \myref{ratwm} for $ma=0.1$, $g=0.1$.
\label{Rorig} }
\end{center}
\end{figure}

\begin{figure}
\begin{center}
\includegraphics[width=3in,height=5in,angle=90]{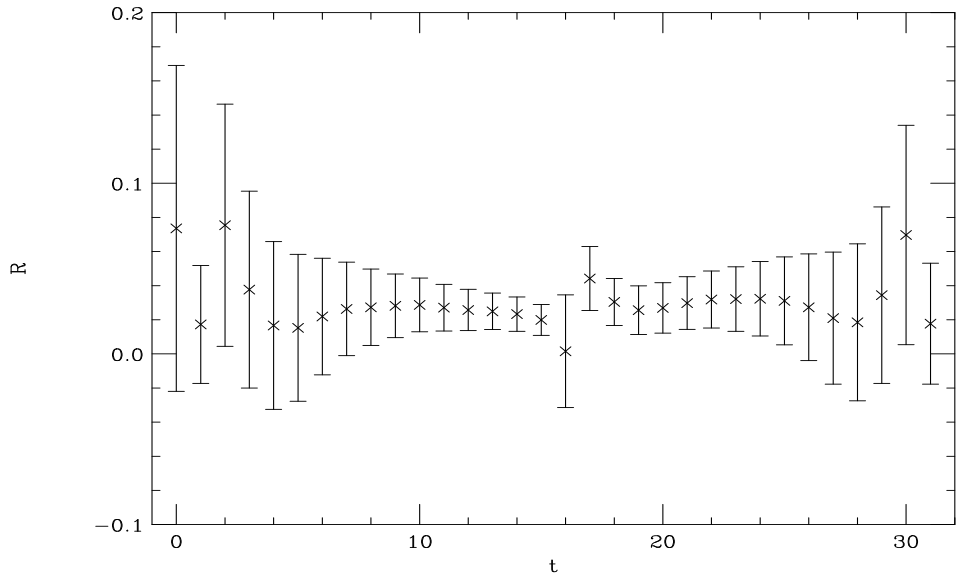}
\caption{The ratio \myref{ratwm} for $ma=0.1$, $g=0.1$, $m_2^2 a^2=0.01$, $\lambda_1 = 0.01$.
\label{Rft} }
\end{center}
\end{figure}

Here we use Monte Carlo simulations to nonperturbatively
measure the ratio \myref{ratwm} using the almost naive supercurrent \myref{latsuc}.
The simulation method is rational hybrid Monte Carlo \cite{Kennedy:1998cu}, and
the runs were performed on Compute Unified Device Architecture (CUDA) enabled graphics processing
units (Nvidia GeForce GTX 285, GTX 480 and Tesla C1060), 
using code that we developed and tested in our previous work.  We have
measured the autocorrelation time to be approximately 12 molecular
dynamics time units for $ma=0.1$ bare fermion mass, with
a coupling $g=0.1$.  The length of the simulation was 5,000 molecular
dynamics time units and we sample at each 5 time units.
Errors in the ratio function $R(t)$ are computed by jackknife
analysis with data blocked into 5 samples each.  The runs
are summarized in Table \ref{oruns}.  As can be seen we also
consider the case of the fine-tuning action \myref{symgeac}
except that we allow for a bare fermion mass $m$ and
set $y_1=g$.  Results for $R(t)$ in each case are displayed
in Figs.~\ref{Rorig} and \ref{Rft}.  It can be seen that at large times,
where the long distance theory should be obtained,
$R(t) \lappeq \ord{g^2}$.  This is consistent with nonconservation
of the almost naive supercurrent beginning at two loops or higher.
It can also be seen that the action \myref{acws} gives a significantly
smaller value for $R(t)$ than the action \myref{symgeac}---again, a fermion
mass term has been added to the latter.  This shows how the
formulation \myref{acws} has a particularly small violation
of supersymmetry, even at higher orders.

\begin{table}
\begin{center}
\begin{tabular}{ccccc}
\hline
$ma$ & $g$ & $m_2^2 a^2$ & $\lambda_1$ & lattice \\ \hline
0.1 & 0.1 & --- & --- & $8^3 \times 32$ \\ 
0.1 & 0.1 & 0.01 & 0.01 & $8^3 \times 32$ \\
\hline
\end{tabular}
\caption{Parameters of the Monte Carlo simulations that
we have performed. \label{oruns} }
\end{center}
\end{table}

\section{Conclusions}
\label{concl}
We found that at one loop $\vev{\p_\mu^S S_\mu(x) \Ocal(0)}=0$.
This was true without any tuning of the lattice action,
and provided the almost naive supercurrent is used. 
The numerical results were explained by the fact that the one-loop diagram
is related to a free theory diagram, and so must vanish.
We showed that this result does not hold if a more naive discretization
is used.  Next we discussed two loop diagrams where we do
not expect the cancellations to hold, since they
are sensitive to the mismatch in self-energies that
was already found in our previous study of one-loop
counterterms.  We look forward to presenting numerical
results for the two loop diagrams in a forthcoming paper.
Finally, we provided nonperturbative results with the
almost naive supercurrent.  It was seen that the
nonconservation of the supercurrent is consistent with
contributions beginning at two loops.  Another
direction for future research is the fine-tuning of the
action together with the search for the renormalized
supercurrent, which will have the more general
form \myref{fpas}.  Investigations in this direction
are in progress.

\section*{Acknowledgements}
This research was supported by the Dept.~of Energy,
Office of Science, Office of High Energy Physics, Grant No.
DE-FG02-08ER41575.

\end{document}